# Role of high nitrogen-vacancy concentration on the photoluminescence and Raman spectra of diamond

*Mona Jani*, Mariusz Mrózek, Anna M. Nowakowska, Patrycja Leszczenko, Wojciech Gawlik, Adam M. Wojciechowski**

M. Jani, M. Mrózek, W. Gawlik, A. M. Wojciechowski
Institute of Physics, Jagiellonian University, Łojasiewicza 11, 30-348 Kraków, Poland
E-mail: mona.jani@uj.edu.pl; a.wojciechowski@uj.edu.pl

A. M. Nowakowska, P. Leszczenko
Faculty of Chemistry, Jagiellonian University, Gronostajowa 2, 30-387 Kraków, Poland

We present a photoluminescence (PL) and Raman spectroscopy study of various diamond samples that have high concentrations of nitrogen-vacancy (NV) color centers up to multiple parts per million (ppm). With green red, and near infrared (NIR) light excitation, we demonstrate that while for samples with a low density of NV centers the signals are primarily dominated by Raman scattering from the diamond lattice, for higher density of NVs we observe a combination of Raman scattering from the diamond lattice and fluorescence from the NV centers, while for the highest NV densities the Raman signals from diamond are completely overwhelmed by the intense NV's fluorescence. However, under NIR excitation, Raman diamond signatures can be observed for some diamonds. These observations reveal different roles of two mechanisms of light emission and contradict the naïve belief that Raman scattering enables complete characterisation of a diamond crystalline sample.

## 1. Introduction

Raman spectroscopy is a popular non-destructive, high-resolution technique, widely used for various material studies. It is particularly useful for the analysis of the peculiarities of collective excitations (e.g. by optical phonons) in carbon materials, especially in diamond.[1–3] The first-order Raman spectrum of a monocrystalline diamond (MCD) consists of the triply degenerate $T_{2g}$ Brillouin zone-center vibration mode of the two interpenetrating face-centered cubic lattices. The vibration energy corresponds to the 1332 $cm^{-1}$ shift of the Raman scattering line relative to the exciting light frequency or wavenumber. Analysis of the detailed shape of the Raman modes, such as their wavenumbers, widths, and intensities provides a valuable insight into the lattice defects, crystallite size, and various effects of external perturbations.[4] The Raman features of nanocrystalline diamond (ND) powders are different from those of MCD.[5] In the shape of the first-order Raman line, the main difference is the asymmetric broadening and peak shift to the lower wavenumbers (within a region of 1322 - 1328 $cm^{-1}$) with decreasing crystallite size.[6] According to Refs.,[7,8] this is caused by the effect of phonon confinement that arises from the crystal imperfections and increasing effects of non-diamond phases that become important when crystal size decreases. The most relevant non-diamond phases are detectable between 1200 $cm^{-1}$ - 1500 $cm^{-1}$ as the disordered carbon (D-band) and graphite-related component (G-band) seen close to 1600 $cm^{-1}$. Analysis of the intensity ratio of the Raman diamond peak and the G-band, as well as the width of the diamond Raman line, enables assessment of the phase purity and crystalline quality of the diamond material.[9] The crystalline diamond constituents are characterized by Raman components that are shifted by a characteristic wavenumber interval (here 1332 $cm^{-1}$) relative to the exciting laser, in contrast to the non-diamond ones whose spectral features shift and change amplitudes when the laser wavelength is changed.[10] In addition to the discussed Raman features in regular diamond structures, various color centers, particularly NVs, play an important role in optical spectra of diamond samples. The NV center is a point lattice defect consisting of a substitutional nitrogen atom with an adjacent carbon vacancy.

The formation of NV color centers in diamond evoked a strong interest in that material due to its spectacular properties, such as strong and stable PL[11] and feasibility of the electronic spin resonance studies with optical detection and long coherence times.[12–14] Bright NV diamonds triggered extensive research in a wide range of applications, e.g. in quantum information processing,[15] sensorics (especially magneto- and thermometry),[16] and intracellular bio-particle detection.[17–19] For practical applications, procuring high densities of NV centers in the diamond crystals and powders is essential. For such purposes, the selection of type Ib high pressure high temperature (HPHT) diamond is ideal due to their natural content of isolated substitutional nitrogen, between 100 and 200 ppm and more.[20] Irradiation of this diamond with high energy particles such as electrons, protons, neutrons or ions generates lattice vacancies that are further recombined by annealing with nitrogen to produce the NV centers.[21,22] The NV centers formed in this procedure are both negatively charged ($NV^-$) and neutral ($NV^0$), with their zero-phonon lines (ZPL) at 575 nm and 637 nm, respectively, followed by a broad phonon luminescence sideband with the highest intensity around 700 nm.[23] The density of $NV^-$ is typically higher than $NV^0$ because substitutional nitrogen acts as an electron donor to the vacancy. The PL spectra of NV-rich diamonds powders are much more sensitive to external influences than those of the regular MCDs,[24] which is important for sensoric applications.

The PL and Raman spectra in diamond are often observed using green laser excitation corresponding to 2.33 eV energy. Absorption of such photons and subsequent Stokes scattering to an excited vibration level positioned about 165 meV above ground state results in optical emission of a wavelength of 572.6 nm that is, in the appearance of the 1332 $cm^{-1}$ Raman line.[25] This is close to the characteristic fluorescence of $NV^0$ and $NV^-$ which have their ZPLs at 575 nm (Raman line at 1406 $cm^{-1}$) and 637 nm (Raman line at 3090 $cm^{-1}$), respectively. Consequently, for the 532 nm laser excitation, the Raman spectra and PL may overlap. This is an important point as it indicates the interplay of two mechanisms. In the following, we present both types of spectra recorded using Raman spectroscopy setup with several different diamond samples and discuss how they depend on diamond properties, particularly on the density of NV centers. Our work is motivated by discussions we had within the NV-diamond community on observability of the Raman signal at high NV densities. Although it is visible for single-NV samples and NV ensembles fabricated as shallow and nanometer-thin layers, we demonstrate that with high-NV density bulk and sub-micron diamonds it is no longer the case. To our knowledge, there are not many reports raising the issues related to the competition of Raman scattering of diamond and PL of dense NV samples in the same spectra.

Below, we present spectra for Ib HPHT ND powders, and fluorescent ND (FND) powders with crystallite size 140 nm and 1 μm having high densities of NV centers. In addition, we used three MCDs, one with a low NV density, and two others that were volume and surface irradiated with electron and proton beams and annealed to yield higher and spatially varying densities of NV centers. Raman and PL spectra collected at 532 nm excitation of the MCD have their amplitudes strongly dependent on the NV concentration. In particular, for high NV densities, the characteristic Raman features become completely obscured by strong NV fluorescence. The observed variations of the Raman scattering and PL amplitudes reveal the existence of two competing mechanisms contributing to the light emission. Depending on the specific NV concentration and excitation wavelength, either a coexistence of two different signatures corresponding to a specific mechanism or a domination of either of them is observed. This contradicts the above-mentioned paradigm that Raman scattering offers universal characterization of diamond samples.

## 2. Experiment section

### 2.1. Sample preparation

In this work we have used three types of commercially purchased sub-micron sized crystals and three bulk diamonds of which two bulk diamond plates were subjected to proton/electron irradiation to enhance the NV concentration.

The first type of NDs were HPHT-type powders, with an average crystallite diameter of ~ 167 nm (measured by dynamic light scattering) purchased from Microdiamant AG. These NDs were suspended in water, sonicated and dried on a glass microslide. The second and third types of NDs were carboxylated FND slurries with NV centers concentration of 3 ppm and average sizes of 140 nm and 1 μm. These FNDs were purchased dispersed in deionized (DI) water (1 mg/ml), from Adámas Nanotechnologies. 40 μl from the slurry was naturally drop-dried on the $CaF_2$ substrate (Crystran LTD, Poole, UK, Raman grade).

For measurements of bulk diamond spectra, we have used three (100)-oriented, polished diamond plates purchased from Element Six: two types of Ib HPHT diamonds (MCD-1 and 2) and one type IIa electronic-grade CVD sample (MCD-3). The MCD-1 sample of 3.12 × 3.10 × 0.60 $mm^3$ size had an initial nitrogen concentration of $[N_i]$ ∼ 380 ppm (measured by EPR). This sample was irradiated with a non-spatially-uniform 3 MeV energy, electron beam and subsequently annealed for 4 h in vacuum at ∼ 750 °C resulting in the creation of a large volume with a lateral gradient of NV concentration that has been characterized in Ref.[26]

The MCD-2 was a plate of 3.0 x 3.0 x 0.3 $mm^3$ with $[N_i]$ ∼ 50 ppm. It was implanted with a proton beam of 1.8 MeV energy from a Van de Graaff accelerator focused on a spot of ~ 20 μm diameter. Subsequent 2 h of annealing in vacuum at ~ 900 $^oC$ resulted in a creation of a high volume of NV concentration close to the diamond surface and that extended to depths of ~ 20 μm, which is the proton stopping range at this energy (as simulated in the SRIM package).[27] The relatively high proton fluence of $4.5\times10^{16}$ $p^+/cm^2$ was responsible for the observed damage in the center of the implanted area and resulted in a non-uniform density profile of NV around the implantation spot.[28] The last bulk sample, MCD-3, has not been processed in any way and had concentrations of nitrogen and NV centers specified by the manufacturer to be [N] < 5 parts per billion (ppb) and [NV] < 0.03 ppb.

### 2.2. Spectroscopic Characterization

Raman spectra in the visible were collected using a WITec Alpha 300 confocal Raman microscope (Ulm, Germany) equipped with the 532 nm and 633 nm excitation wavelengths and a CCD detector (Andor Technology Ltd, Belfast, Northern Ireland). A 20x air objective was used (Zeiss EC Epiplan 20x, NA=0.4, Oberkochen, Germany). The spectral resolution was 3 $cm^{-1}$. The applied laser powers were 1.9 mW for the 532 nm laser and 15 mW for the 633 nm one. For MCD measurements, the 532 nm laser power was reduced to 0.185 mW. The resulting spectra were obtained with 10 accumulations and an acquisition time of 0.5 s for both 532 and 633 nm lasers. FNDs were deposited from the suspension in DI water on $CaF_2$ slides or glass microslides and left to dry. For each sample, at least 10 spectra were collected at different locations.

For the NIR excitation, an FT-Raman spectrometer (Bruker MultiRAM) was used to generate Raman spectra of the samples in the spectral range of 4000-400 $cm^{-1}$, with a spectral resolution of 4 $cm^{-1}$, set to 32 scans per sample. The spectrometer was equipped with a diode-pumped Nd:YAG laser set at 300 mW power, with excitation wavelength of 1064 nm, a germanium detector cooled with liquid nitrogen, and the instrument was controlled with OPUS 7.0 software.

## 3. Results and discussion

### 3.1. Raman and photluminescence spectra of pristine ND and FND

The Raman spectra of both pristine ND and FNDs excited by visible and NIR lasers at a wavelength of 532 nm, 632 nm, and 1064 nm are shown in **Figure 1a,** 1b, and 1c, respectively. Measurements were performed at 10 arbitrarily chosen locations for each sample and the data are plotted for a single, representative one. Due to substantial variations in the diamond-film thickness and different numbers of contributing particles we could not compare the signal amplitude for various samples. Still, significant characteristic shape changes between the spectra could be observed and are discussed in the following.

For pristine NDs, the only visible peak at ~ 1330 $cm^{-1}$ is the first-order excitation of the triply degenerate ($T_{2g}$) optical phonon at the center of the Brillouin zone. A broad feature at 1450 - 1800 $cm^{-1}$ is commonly attributed to the graphitic carbon (G-band). The HPHT diamonds have a highly uniform crystal structure with a low concentration of dislocations or lattice twinnings. Therefore, the presence of G-band is a consequence of the crushing and milling processes used for the preparation of small NDs from the larger synthesized diamond crystals. Non-diamond carbon phases can, in principle, be removed from the ND surface by oxygen treatment or acid cleaning, but this has not been done for the samples we have acquired. The ND spectra show a distinct diamond Raman peak and a negligible PL contribution for visible excitation (Figure 1a and 1b). The situation changes at NIR excitation, where the diamond Raman peak is still visible, however, a noticeable background becomes present.

The FNDs procured with crystallite sizes of 1 μm and 140 nm contain NVs in both neutral ($NV^0$) and negative ($NV^-$) charge states. The neutral center, $NV^0$ has the ZPL at 575 nm (2.156 eV) associated with the transition between the excited $^2A_2$ and ground $^2E$ states. It can be detected using optical excitation in a broad range of wavelengths from 400 to 575 nm and has emission in the wavelength range from 575 to 700 nm. In case of $NV^-$ center, the ZPL transition at 637 nm (1.945 eV) occurs between the $^3A_2$ ground state and an excited $^3E$ triplet. The optical excitation spectrum stretches from 450 nm to 650 nm, while the PL emission range spans from around 600 nm to about 800 nm. Specifically, the Raman and PL spectra of FND may overlap. Consequently, for sufficiently high concentration of NVs, the FND spectra in Figure 1a are dominated by fluorescence with no visible diamond Raman contribution.

Figure 1b presents emission spectra of the FNDs excited with the 633 nm wavelength laser. With red excitation, the emission in the phonon sidebands is restricted to $NV^-$ centers only, as the excitation wavelength is too long for $NV^0$. The sloped background seen in Figure 1b represents the tail of the phonon sideband of the strong $^3E \rightarrow {}^3A_2$ transition.[29] Here, also, no Raman peak at 1330 $cm^{-1}$ is visible for the FNDs, while it remains the dominant feature for the pristine NDs.

To complement the information obtained by analysing the Raman spectra for FNDs where the peaks corresponding to NV centers ($NV^0$ and $NV^-$) are observed, we collected spectra in the NIR excitation and detection region with a focus on the visibility of the diamond peak. Figure 1c shows the Raman and PL spectra for ND and FND collected using 1064 nm laser excitation.

For pristine NDs, the obtained NIR Raman spectrum shows a weaker diamond peak at ~ 1330 $cm^{-1}$ and broad fluorescence with a maximum around ~ 2800 $cm^{-1}$ that is expected to originate from non-diamond carbon, as the NDs were used without purification. As shown in Refs.[30–32], Raman scattering intensities strongly depend on the nature of the scattering phase and on the excitation wavelengths. Wagner *et al.*[33] showed that the Raman intensities of diamond decrease and those of non-diamond carbon increase when the excitation wavelength is shifted from the visible to the infrared range. Small amounts of non-diamond carbon phases, which cannot be detected with visible Raman scattering, may be clearly seen with IR-Raman spectroscopy. This non-diamond carbon is removed before creating NV centers in the FNDs by an annealing and oxidation method, and is not detectable in NIR Raman spectra. The Raman spectrum for 1 μm FNDs exhibits an intense diamond peak at 1330 $cm^{-1}$ with a background fluorescence tail from the NVs. The observed diamond peak proves that in 1 μm FNDs the Raman scattering vibrations are due to the diamond phase of the intact diamond lattice. On the other hand, we observe a noticeable difference from the spectrum of 140 nm FNDs where the diamond peak is not observed. Similar spectra to those of 140-nm FNDs were observed also for FNDs with selected sizes below 140 nm, e.g. (100 nm, 70 nm, 60 nm, 50 nm, 40 nm, 30 nm, and 20 nm), purchased from Adámas Nanotechnologies (*not shown here*) and they also did not demonstrate the characteristic Raman peak.

### 3.2. Raman spectra of MCDs with varying NV concentrations

#### 3.2.1. Volume irradiated diamond

The MCD-1 sample has a non-uniform NV concentration with a strong gradient, varying from 4 to 36 ppm over a distance of just 1.5 mm, i.e. ~ 21 ppm/millimetre. Its optical microscope image is shown in **Figure 2a**. The NV centers were created inhomogenously due to the uneven profile of the electron beam used for irradiation.[26] The sample looks yellow on the outer sides, as is typical for the type Ib HPHT diamonds, and the color gradually changes to dark brown toward the center of the electron-irradiated area. Most importantly, electrons at 3 MeV have a mean-free path of several millimetres in diamond, resulting in a uniform creation of vacancies throughout the depth (thickness) of the diamond plate. The radiation damage caused to the diamond lattice is, therefore, relatively uniform throughout the depth of the diamond, and changes only laterally due to the electron beam profile. In particular, this process ensures a high quality of the diamond-lattice near the surface.

Figure 2b shows the Raman and PL spectra of MCD-1 collected along the straight line, at the marked positions shown in Figure 2a using green laser excitation at 532 nm. The spectra are corrected for the baseline for better visualization of the results. The NVs concentration gradually changes from the lowest, at position $x$ = 21 μm, to the highest around $x$ = 1286 μm. As expected, typical diamond Raman signals are observed at 1332 $cm^{-1}$, along with the ZPL signals of $NV^0$ at 1430 $cm^{-1}$ / 576 nm and $NV^-$ at 3131 $cm^{-1}$ / 637 nm. In the inset, a zoomed-in view of the diamond peak in addition to the PL signals of the $NV^0$ centers is presented and shows that the intensity of the diamond peak gradually decreases as the intensity of the $NV^0$ peak increases with increasing position coordinate. Ultimately, the diamond Raman peak disappears for the highest NV concentrations because of the overwhelming $NV^0$ PL. To quantify this observation, multiple spectra were collected along the marked line (Figure 2a), and the characteristic peak amplitudes are shown in Figure 2c. From the figure, it is evident that the peak intensities of $NV^0$ and $NV^-$ are gradually increasing toward the irradiated area centre, that is, with increasing concentration of NVs. The PL of the $NV^-$ centers has a higher intensity being two orders of magnitude larger than that of $NV^0$ in the non-irradiated region and more than an order of magnitude larger in the irradiated area. However, the diamond Raman peak amplitude remains almost constant throughout the positions where it is visible. The horizontal dashed line in Figure 2c demonstrates the independence of the mean detected amplitude of the Raman peak on the NV concentration. It allows comparing the ratio of $NV^0$ ZPL and diamond peak amplitudes and it can be seen that until a certain concentration of NV centers is reached (i.e. the window between $x$ = 500 - 800 μm) the diamond peak is detectable and later on it is conquered by the overwhelming intensity of $NV^0$ PL. The stable amplitude of the diamond peak over the entire range where it is detected suggests that the diamond peak is not diminishing, at least in the regions where the diamond is not heavily radiation-damaged.

### 3.2.2. Surface-irradiated diamond

To verify the observations of the effect of NV concentration on the Raman diamond peak presented in Sec. 3.2.1, we have also studied a proton-implanted sample, MCD-2, which had vacancies created by the $p^+$ beam in the top 20-μm-thin layer. Due to subsequent annealing and stimulated vacancy migration, this sample exhibits NV concentration gradients on the micrometer scale around the implanted spot. Its fluorescence image with green excitation is shown in **Figure 3a**. The implanted area shows high fluorescence on a nearly-zero background. It has an irregular shape, which is due to the focussing and instability of the proton beam, and in the centre exhibits a drop of fluorescence, which is caused by the radiation damage. Optical spectra have been recorded along the indicated green line, and several spots have been marked in the region where the NV concentration follows a clear spatial gradient with point 1 marking the lowest NV concentration (darkest area) and successive point numbers indicating increasing NV fluorescence.

Figure 3b depicts the Raman and PL spectra collected with green excitation at 532 nm in the marked positions on the sample. As expected, at point 1, we observed signals from the characteristic diamond Raman peak at wavenumber 1332 $cm^{-1}$, $NV^0$ ZPL at wavenumber 1430 $cm^{-1}$ (~ 575 nm) and $NV^-$ ZPL at wavenumber 3131 $cm^{-1}$ (~ 637 nm). In the inset of Figure 3b, a zoomed-in view of the diamond peak is shown in addition to the ZPL line of $NV^0$. It is apparent that the diamond peak visibility is again affected by the $NV^0$ fluorescence at higher NV centers concentrations and eventually becomes undetectable. To verify whether this is an indication that the Raman diamond peak is decreasing or covered by a strong fluorescence from the NV center, fitted amplitudes of the characteristic spectral features of diamond, $NV^0$, and $NV^-$ are presented in Figure 3c. Here, the horizontal broken line again demonstrates that the diamond peak amplitude remains stable when moving towards the center of the implantation region (~ 90 μm), while the density of the NV centers gradually increases. Only in a relatively narrow area around the center is the $NV^-$ fluorescence dropping due to a significant damage to the diamond caused by a high implantation dose. This is evident in Figure 3a, which shows that the damaged area as a black spot within the fluorescent region. Although the concentration of $NV^0$ is maintained at a relatively stable level in this area, the imaging system used for take this image was spectrally sensitive only to the $NV^-$ PL. The diamond peak for this sample becomes hindered in the range of positions from ~ 25 μm to 120 μm. Outside of this area it maintains a stable value, and again we can compare it to NV signals. Both ZPLs have significantly larger amplitudes, and the nearby $NV^0$ ZPL can reach orders of magnitude larger values and bury the Raman signal. Since the maximum NV concentration depends on the initial nitrogen content of the sample, $N_i$, the maximum $NV^0$ ZPL to diamond Raman peak amplitude ratio is also a function of [$N_i$] and should be lower for MCD-2 compared to MCD-1. This is supported by the slightly higher values of the maximum ZPL to Raman amplitude ratio in Figures 2c than in Figure 3c. Therefore, we conclude that such ratios can be calibrated to provide information on the NV density in other dense samples ([NV] ~ 0.1-10 ppm) wherever the thickness of the NV layer is uniform in the volume probed by the microscope. Therefore, combined PL and Raman spectroscopy can, become an essential diagnostic tool to collect information on the creation of NV centers in dense ensemble samples.

### 3.3. NIR Raman spectra of MCDs

Shifting the excitation wavelength in Raman spectroscopy to a region where no PL could be excited is a well-known method of studying undistorted Raman spectra. We applied this approach to study our diamond samples with a NIR excitation of 1064 nm wavelength, i.e. double that of green lasers typically used for NV-diamond experiments. Large detuning from the sample's absorption band enables elimination of the overlap of PL and Raman spectra but, since the Raman scattering efficiency scales with the light wavelength as $\lambda^4$, the NIR excitation results also in a considerable suppression of Raman signals. **Figure 4a** shows the obtained NIR Raman spectra from sample MCD-3 that has the lowest concentrations of nitrogen and NVs. As expected for this ultra-pure diamond, the spectrum contains only the sharp diamond line at wavenumber 1332 $cm^{-1}$ with no visible fluorescence background. On the other hand, NIR-Raman spectra collected at four locations on the MCD-1 gradient sample (Figure 4b**)** shows some fluorescence background in addition to the Raman diamond line at 1332 $cm^{-1}$. The inset shows that the intensity and width of the diamond peak hardly changes between the spots and the Raman signal consistently remains well pronounced on the PL background. Thus, the 1064 nm laser excitation appears as a convenient alternative for measuring the undistorted Raman signals of diamond and avoiding major contributions from NV centers. We note, however, that for long wavelengths and small particle sizes, on the order of 100 nm or smaller, the Raman signals are very weak (Figure 1c) and collecting the spectra can be pretty challenging. Also, in case of diamonds containing other color centers which fluoresce in the NIR band, excitation at 1064 nm may not be a sufficient remedy.

## 4. Conclusions

We have presented optical spectra of bulk and sub-micrometer diamonds under several excitation wavelengths. Under the most common, green light excitation, diamond Raman and NV PL spectra, signals that can be associated with two different physical mechanisms of light scattering, are spectrally overlapping and, therefore, not easily discernible. Moreover, in case of high-NV-density samples ([NV] > 1 ppm), the $NV^0$ PL can partially or even fully obscure the diamond Raman line. Through Figures 2 and 3, we have experimentally demonstrated that the diamond peak in the optical spectra is not detectable when the NV centers concentration become high and is hidden under the strong fluorescence arising from $NV^0$ centers. In fact, this observation holds not only for the green excitation, but extends to almost entire visible range, as the NV centers's PL can be excited from the blue ($NV^0$) up to the red ($NV^-$) wavelengths.

We hope these observations shed some light on and help explain the reasons for the absence of Raman signal in NV-rich diamond samples, which sometimes triggers discussions in the NV-diamond community. Our observations also pave the way to establishing a new method for calibration of a high NV-density by comparing the $NV^0$ ZPL amplitude with that of the diamond Raman line. With standardized measurement conditions (light intensity, illuminated area, and the microscope-objective numerical aperture), it should be feasible to calibrate the ratio of $NV^0$ ZPL to diamond Raman peak amplitudes as a function of the NV concentration. This would provide a valuable tool for the characterization of NV-rich samples in the regime where calculations based on scaling the single-NV fluorescence level (photon count rate) becomes inappropriate, e.g. due to the light reabsorption.

While the availability of Raman microscopy setups operating in NIR might be limited (need of dedicated lasers and IR detectors), the efforts of using NIR excitation pay off by avoiding interpretation problems of overlapping spectra.

## Acknowledgements

This work was supported by the Foundation for Polish Science TEAM-NET Programme co-financed by the EU under the European Regional Development Fund, grant no. POIR.04.04.00-00-1644/18.

## Conflict of Interest

The authors declare no conflict of interest.

## Data Availability Statement

The data that support the findings of this study are available from the corresponding author upon reasonable request.

Figures

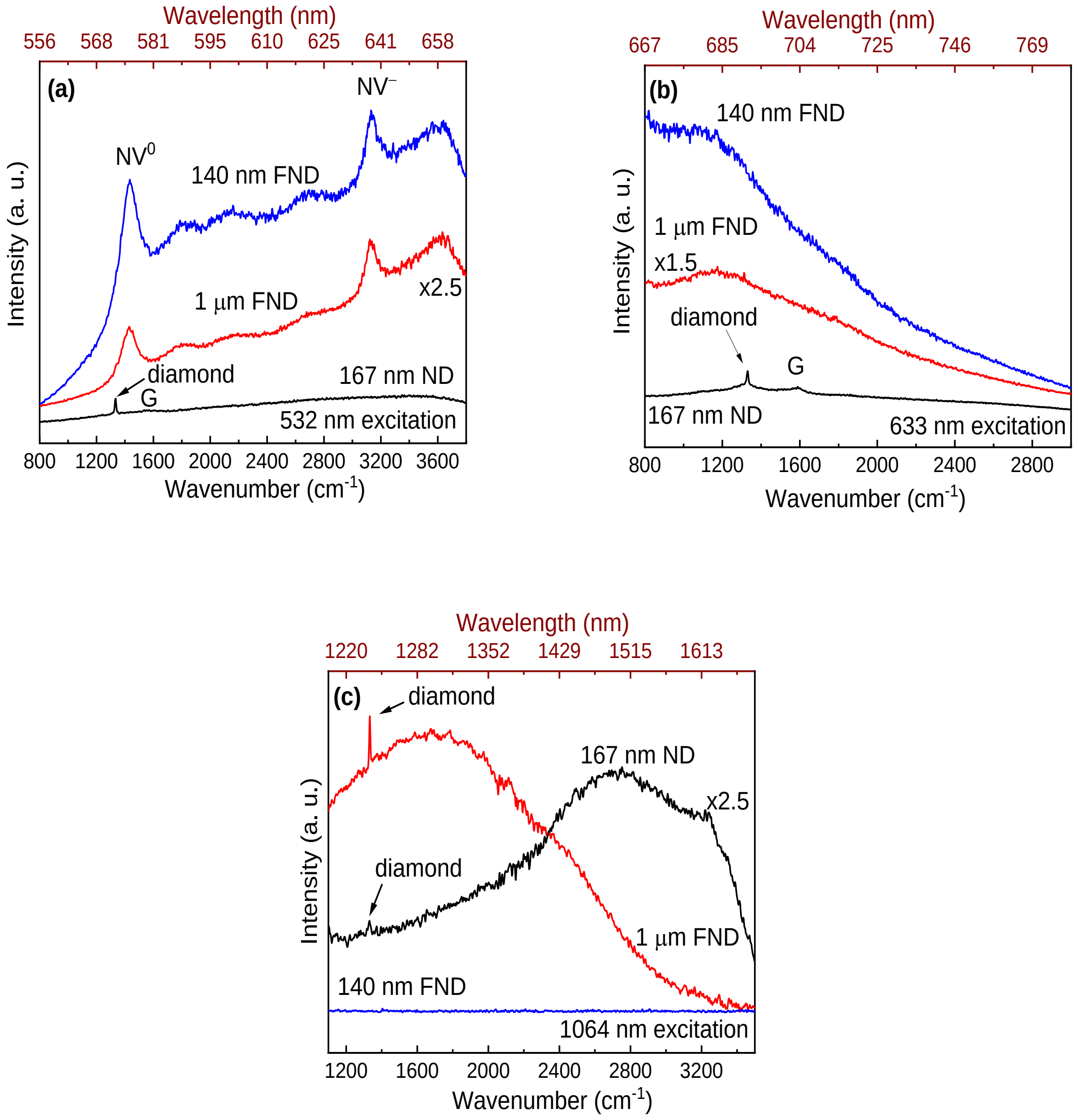

**Figure 1.** PL and Raman spectra for ~ 167 nm HPHT ND, 1 μm and 140 nm FND powders collected with: 532 nm (a); 633 nm (b); and 1064 nm (c) laser excitations. At 532 nm for FNDs we observe PL from $NV^0$ and $NV^-$. Interestingly, we did not observe the diamond Raman peak for FNDs with the 532 nm and observe phonon sideband tail from $NV^-$ at 633 nm. For the pristine ND sample, diamond Raman and graphite (G) peaks are well visible and no resonance fluorescence. At 1064 nm, since NV center fluorescence is very weak in the NIR region, we observe diamond peak for the 1 μm

FND and pristine ND samples. For the 140 nm FND, the Raman signals are very weak and collecting the spectra can be pretty challenging.

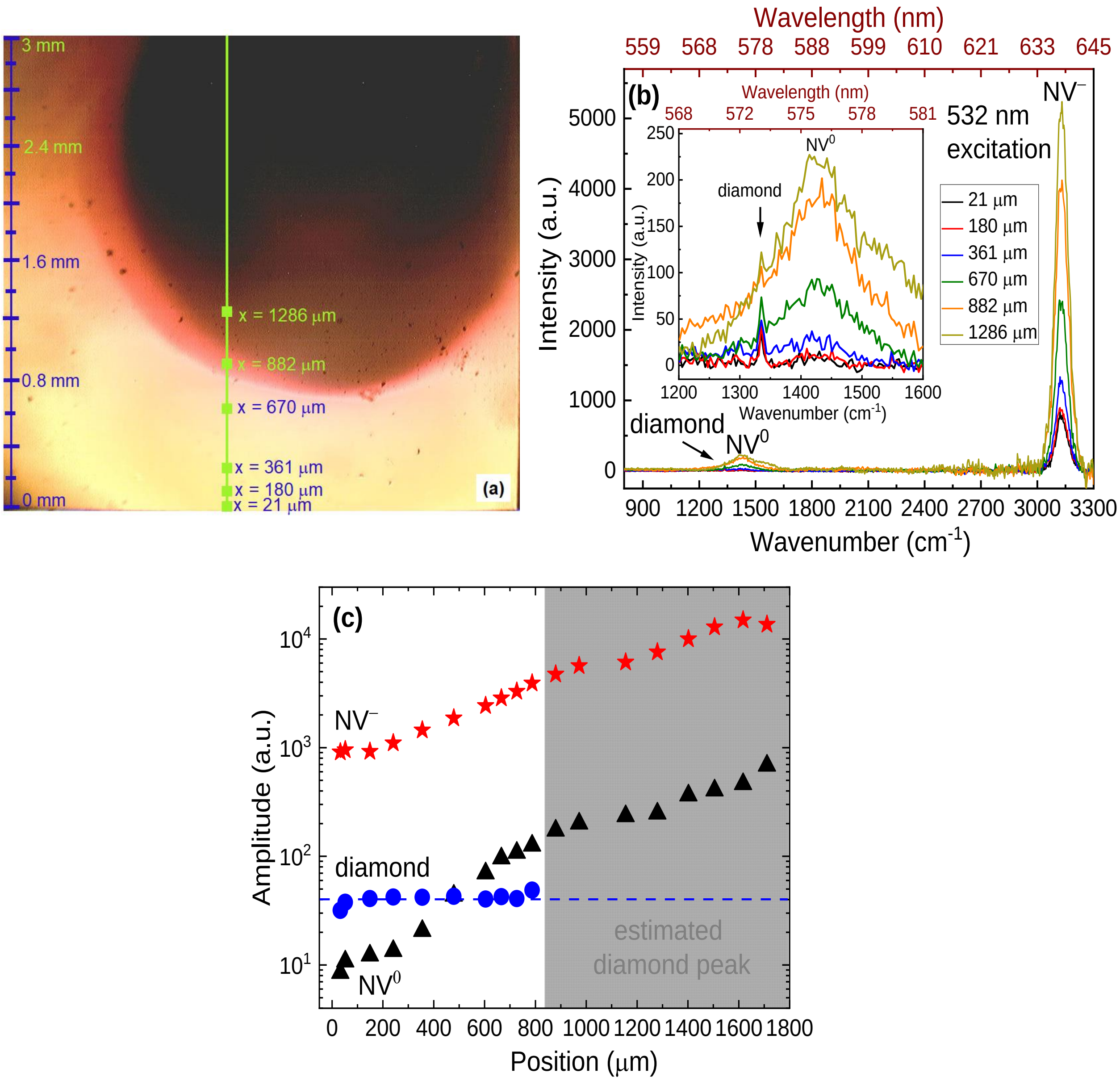

**Figure 2.** a) Image of the MCD-1 sample with a gradual increase in NV density (bright to dark area). b) intensity of the PL and Raman spectra collected at marked *x*-positions along the green line in figure (*a)* with the 532 nm laser excitation. The inset shows a zoom image of the Raman diamond peak (next to the $NV^0$ PL). The inset clearly demonstrates that the relative amplitude of the diamond peak intensity reduces under increasing $NV^0$ fluorescence intensity. c) Amplitudes of the $NV^–$, $NV^0$, and Raman diamond peak for various positions along the green line in image (*a)*. Dashed line demonstrates the independence of the Raman amplitude on NV concentration (for x

= 800 μm the Raman peak is invisible, so the broken line shows only the extrapolated values)

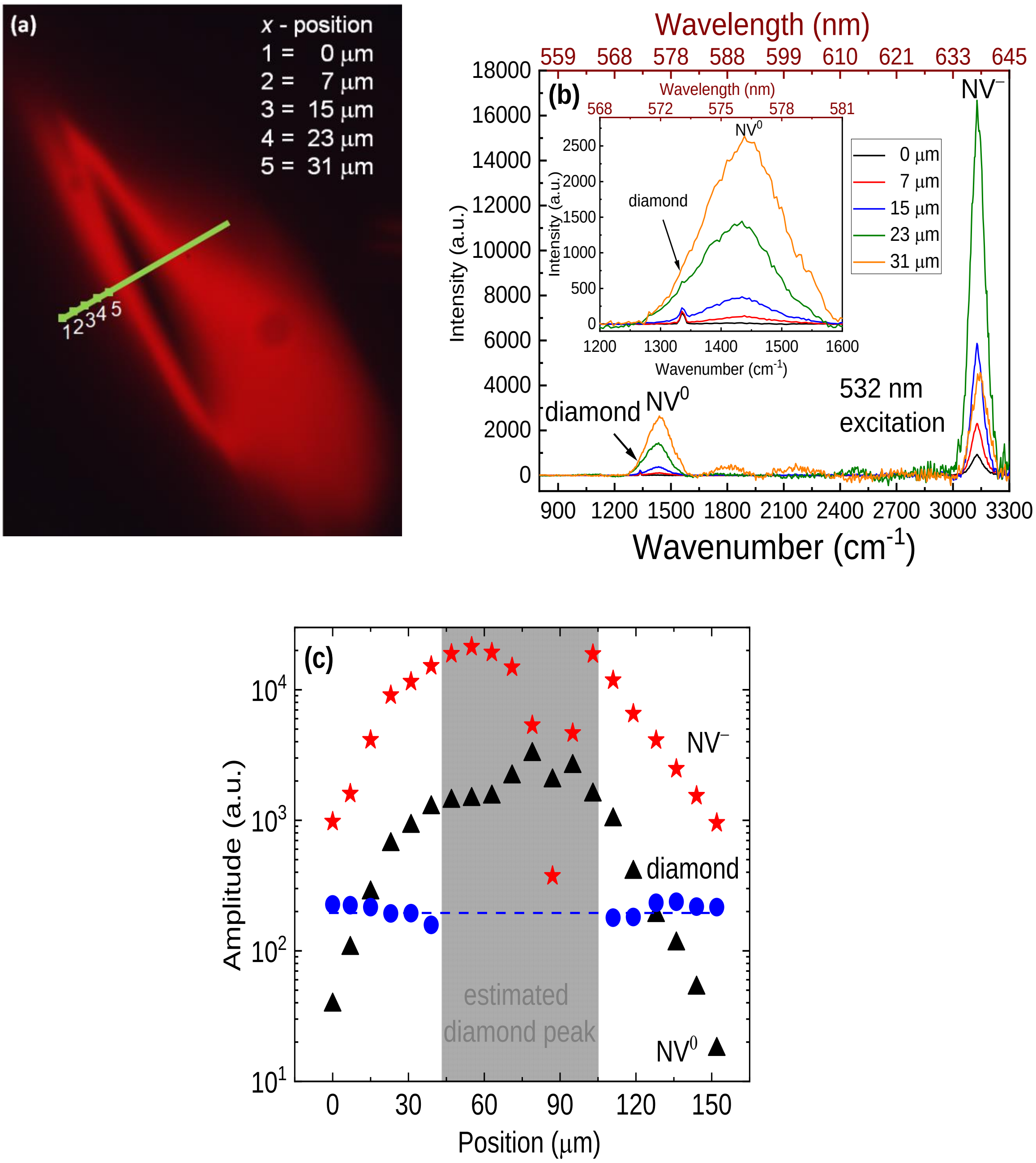

**Figure 3.** a) Image of MCD-2 with varying NV centers concentration (dark to bright red area); b) PL and Raman spectra collected at indicated *x*-positions along green line in *a)* with 532 nm laser excitation. The inset shows zoom image of the diamond peak besides $NV^0$ ZPL line. The inset shows the diamond peak intensity diminishes under $NV^0$ PL intensity; c) Raman peak intensity amplitudes of diamond peak, $NV^-$ and $NV^0$ with respect to the positions along green line in image *a)*. Dashed line is the estimation as diamond peak diminishes completely. $NV^-$ and $NV^0$ peak amplitudes are increasing gradually, and diamond peak although being present is overwhelmed by fluorescence

at certain NV density. The drop in $NV^-$ and $NV^0$ at ~ 90 μm, is as the NV centers are damaged (black spot in center of image *a*).

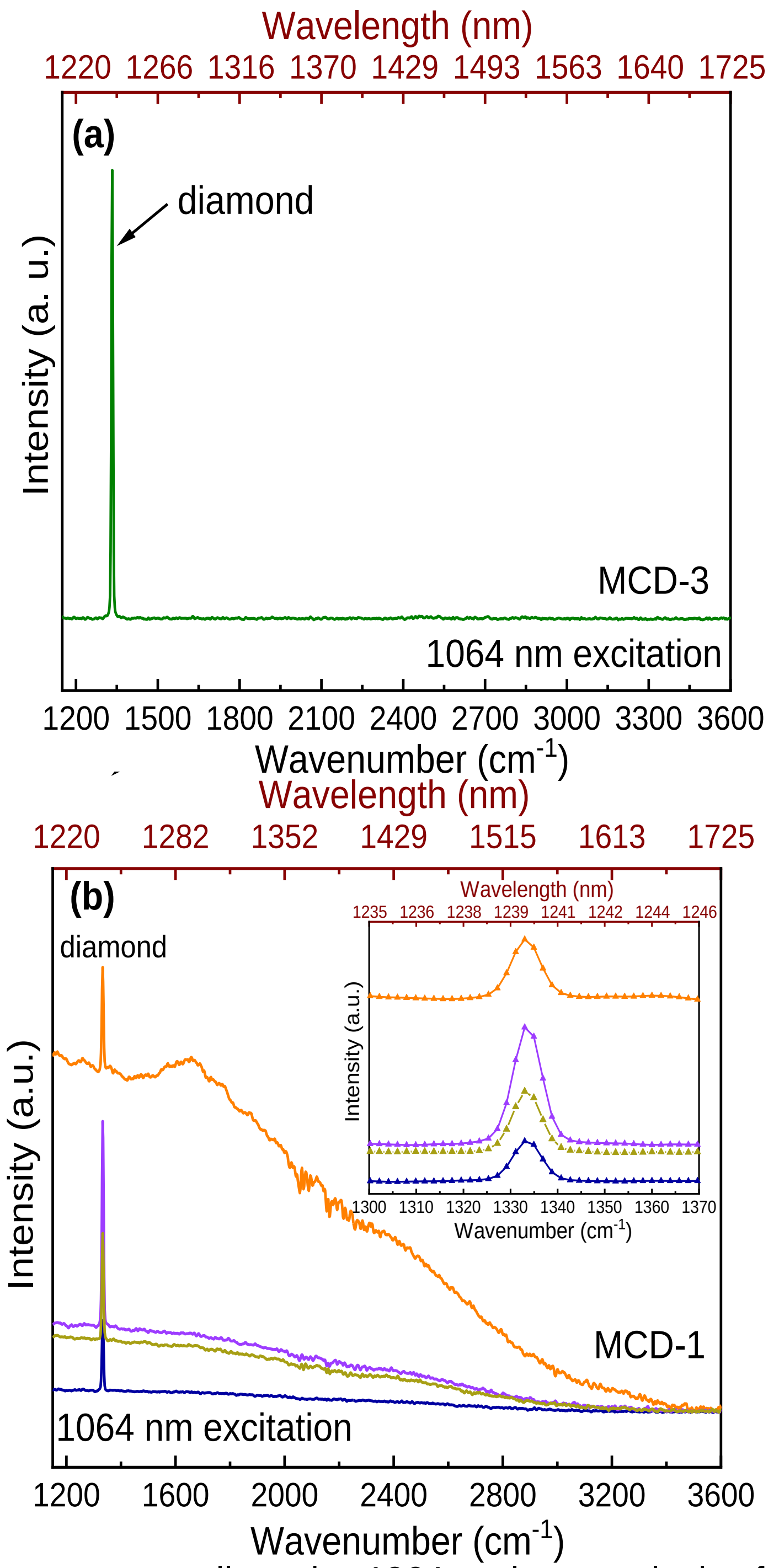

**Figure 4**. NIR Raman spectra collected at 1064 nm laser excitation for a) MCD-3; and b) MCD-1 measured at random locations (Figure 2a). Inset shows the zoom-in signals of Raman diamond peak intensity and width from the corresponding spectra. MCD-3 shows a sharp Raman spectral line at 1332 $cm^{-1}$ with no background fluorescence. For MCD-1, the graph presents a diamond Raman line with some fluorescence

background. The intensity and width of the diamond peak hardly changes between the spots and the Raman signal consistently remains well pronounced on the PL background (seen in inset).